\documentclass[journal]{IEEEtran}
\usepackage{amsmath,amssymb,amsfonts}
\usepackage{algorithmic}
\usepackage{array}
\usepackage[caption=false,font=normalsize,labelfont=sf,textfont=sf]{subfig}
\usepackage{textcomp}
\usepackage{stfloats}
\usepackage{url}
\usepackage{verbatim}
\usepackage{graphicx}
\usepackage{cite}
\def\BibTeX{{\rm B\kern-.05em{\sc i\kern-.025em b}\kern-.08em
    T\kern-.1667em\lower.7ex\hbox{E}\kern-.125emX}}
\usepackage{balance}
\usepackage[linesnumbered,ruled]{algorithm2e}
\usepackage{color, amsthm}
\usepackage{bm}
\usepackage{booktabs}
\usepackage{multirow}
\usepackage{makecell}

\newtheorem{definition}{\bf Definition}

\begin{document}

\title{An Index Policy Based on Sarsa and Q-learning for Heterogeneous Smart Target Tracking}
\author{Yuhang Hao, Zengfu Wang, Jing Fu, and Quan Pan
\thanks{This work was in part supported by the National Natural Science Foundation of China~(grant no. U21B2008).}
\thanks{Yuhang Hao, Zengfu Wang, and Quan Pan are with the School of Automation, Northwestern Polytechnical University, and the Key Laboratory of Information Fusion Technology, Ministry of Education, Xi'an, Shaanxi, 710072, China.
Jing Fu is with the School of Engineering, RMIT University, Melbourne, VIC, 3000, Australia.
E-mail: (yuhanghao@mail.nwpu.edu.cn; wangzengfu@nwpu.edu.cn; jing.fu@rmit.edu.au; quanpan@nwpu.edu.cn).
(Corresponding author: Zengfu Wang.)
}
}
\maketitle

\begin{abstract}
In solving the non-myopic radar scheduling for multiple smart target tracking within an active and passive radar network, we need to consider both short-term enhanced tracking performance and a higher probability of target maneuvering in the future with active tracking.
Acquiring the long-term tracking performance while scheduling the beam resources of active and passive radars poses a challenge.
To address this challenge, we model this problem as a Markov decision process consisting of parallel restless bandit processes.
Each bandit process is associated with a smart target, of which the estimation state evolves according to different discrete dynamic models for different actions - whether or not the target is being tracked. The discrete state is defined by the dynamic mode.
The problem exhibits the curse of dimensionality, where optimal solutions are in general intractable. We resort to heuristics through the famous restless multi-armed bandit techniques. It follows with efficient scheduling policies based on the indices that are real numbers representing the marginal rewards of taking different actions. For the inevitable practical case with unknown transition matrices, we propose a new method that utilizes the forward Sarsa and backward Q-learning to approximate the indices through adapting the state-action value functions, or equivalently the Q-functions, and propose a new policy, namely ISQ, aiming to maximize the long-term tracking rewards. Numerical results demonstrate that the proposed ISQ policy outperforms conventional Q-learning-based methods and rapidly converges to the well-known Whittle index policy with revealed state transition models, which is considered the benchmark.
\end{abstract}

\begin{IEEEkeywords}
Target tracking, radar scheduling, index policy, Sarsa, Q-learning.
\end{IEEEkeywords}

\section{Introduction}\label{sec:introduction}
\IEEEPARstart{A}{n} active-and-passive radar network (APRN) can make the most of cooperative tracking of active and passive radars with completed multiple target tracking (MTT) tasks~\cite{wang2020joint,gao2017joint,yan2022radar,yan2020optimal,yan2020collaborative}.
Radar scheduling for tracking multiple smart targets, which possess the ability to be aware whether they are being tracked and subsequently adapt their dynamics to better hide themselves, has become important in the past years~\cite{kreucher2006adaptive,savage2009sensor}.
In an APRN, the active radars aim to mitigate the tracking error of smart targets and achieve better tracking precision, while the passive radars are used to reduce the probability of interception \cite{zhang2021low,dai2022composed} and avoid smart targets from maneuvering or hiding.
It follows with a significant trade-off between the pursuit of enhanced tracking performance through active radars
and a low probability of being aware by the smart targets under the premise of covert detections through passive radars.
The challenges lie in the aspect of efficiently allocating the beam resources of a large number of active and passive radars with maximized long-term MTT performance.

We consider the smart target tracking problem by scheduling the beam resources in an APRN, which is typically formulated as a partially observable Markov decision process (POMDP)~\cite{kreucher2006adaptive,pang2019sensor,gongguo2019non,shan2020non}.
Achieving optimality of the Markov decision process (MDP) and/or POMDP manifests difficulties in two points: the large-scale state space and multi-decision combinations in MTT.
The difficulties result in exponential computational demands in the worst case, rendering classical dynamic programming approaches ineffective.
For the restless multi-armed bandit (RMAB) problem \cite{ayesta2022reinforcement,nino2023markovian}, the Whittle index policy was developed through the Whittle relaxation technique with achieved near-optimality~\cite{whittle1988restless, weber1990index,fu2022restless}.
In \cite{ayesta2022reinforcement,nino2023markovian}, each target was associated with a \emph{bandit process} of the RMAB model and was prioritized by the Whittle index policy based on state-dependent real numbers, referred to as the \emph{Whittle indices}.
The Whittle indices were computed independently for different bandit processes, and hence the Whittle index policy exhibits significantly reduced computational cost.
Whittle~\cite{whittle1988restless} conjectured that the Whittle index policy approaches optimality when the number of bandit processes tends to infinity.
It was subsequently proved by
Weber and Weiss~\cite{weber1990index} under the \emph{Whittle indexability} and a non-trivial condition related to the existence of a global attractor of the underlying stochastic process.
In \cite{fu2022restless}, the global attractor was proved to exist in a range of cases.
Since the initial work in \cite{whittle1988restless}, the Whittle index policy has drawn broad attention in the realm of resource scheduling for MTT~\cite{howard2004optimal,nino2011sensor,nino2016whittle,nino2020verification,dance2015kalman}.

Unlike conventional MTT problems assuming full knowledge of the underlying bandit processes~\cite{howard2004optimal,nino2011sensor,nino2016whittle,nino2020verification,dance2015kalman}, in an APRN, the transition matrices associated with smart targets are not known a prior, because of the unpredictable dynamics of the smart targets.
The heterogeneity of the targets will further complicate the analysis of our problem.
Conventional techniques are not able to be applied here.

For RMAB problems with unknown transition matrices, reinforcement learning (RL)-based index policies have been explored in recent years~\cite{fu2019towards,avrachenkov2022whittle}.
RL-based index policies learn and approximate the (Whittle) indices, which quantify the priorities of the corresponding bandit processes, through the state-action/Q functions.
Compared with classical RL methods~\cite{meng2021deep}, the approaches in~\cite{fu2019towards,borkar2018reinforcement,avrachenkov2022whittle} avoid the exponential complexity arising from multi-dimensional decisions and low convergence rate for the learning process by incorporating the RMAB technique.
In \cite{fu2019towards}, a policy referred to as Q-learning the Whittle Index Controller (QWIC) was proposed to approximate Q values, which are functions of discrete states, actions, and the Lagrangian multiplier.
The Whittle index was estimated by minimizing the difference between the Q values under action 1 and action 0.
In \cite{borkar2018reinforcement}, an RL algorithm was proposed and analyzed for web crawler scheduling, which was modeled in a class of indexable restless bandits.
The Whittle indices were derived through linear function approximation.
In \cite{avrachenkov2022whittle}, the relative value iteration (RVI) Q-learning algorithm was utilized to learn the Q values on a faster timescale for a static index, while, the Whittle indices were estimated through a slower time scale.
The authors proposed the Whittle index policy based on such two-time scale learning iterations to solve the restless bandits with average reward.
Similar to \cite{avrachenkov2022whittle}, empirical results demonstrated the efficiency of the Q-learning with Whittle Index (QWI) policy in maximizing the discounted cumulative reward with infinite horizon~\cite{robledo2022qwi}.
In \cite{robledo2022tabular}, the deep Q-network technique was adopted to learn the Whittle indices for the Whittle index policy.
In \cite{wang2023optimistic}, incorporated with the upper confidence bound (UCB1) approach, the UCWhittle algorithm was proposed to solve RMAB with unknown state transition matrices.
A bilinear program was exploited to compute the Whittle indices through a confidence bound in the transition dynamics.
Similar policies based on UCB1 and the Whittle index policy were discussed in \cite{xiong2021learning}.
However, the proofs of convergence in \cite{robledo2022qwi,avrachenkov2022whittle} required that all the bandit processes are homogeneous, which is not applicable to our target tracking problem.
In the past work mentioned above,
the implementation of the (Whittle) index policy with tentatively learned indices/Q values will reversely affect the convergence of the learning process in the future.

In \cite{gibson2021novel}, a novel Q-learning-based index policy was proposed for heterogeneous bandit processes, each of which was associated with two sets of Q values - one for reward and the other for resource consumption.
It follows with a Whittle index policy where the indices were estimated as the instantaneous marginal productivity rate.
In \cite{nakhleh2021neurwin}, a Neural Whittle Index Network (NeurWIN) policy was proposed, where the Whittle indices were learned through deep RL with convoluted transition kernels.
In \cite{biswas2021learn}, a public health problem was formulated as an MDP with discrete ``behavioral" states and analyzed through the Whittle index based Q-learning (WIQL).
Unlike~\cite{fu2019towards, avrachenkov2022whittle, robledo2022qwi, gibson2021novel}, the indices were approximated by the difference between the learned Q values for active and passive actions.
In \cite{biswas2021learn}, the WIQL was proved to possess the same optimality guarantee as the traditional Q-learning algorithm.
Nonetheless, the convergence rate of WIQL is slow in the early learning stage.
Building on the state-action-reward-state-action (Sarsa) algorithm~\cite{sutton2018reinforcement}, in~\cite{wang2013backward}, the backward Q-learning based Sarsa algorithm (BQSA) was proposed that integrated the advantages of Q-learning and the Sarsa algorithms.

Applying the Q-learning-based index policies for the multi-smart-target tracking problem presents challenges.
The low convergence rate makes the Q-learning-based methods less appropriate for time-sensitive MTT scenarios, such as the APRN, where inappropriate tracking strategies may fail to capture the dynamics of the smart targets.
The heterogeneity of the targets makes the memory-sharing scheme of action-state pairs in \cite{avrachenkov2022whittle} infeasible, which further complicates the Q-learning process since each target needs to learn separately.

We consider the online radar scheduling for tracking heterogeneous smart targets with distinct and unknown state transition models in an APRN.
We formulate the problem as an RMAB consisting of bandit processes associated with the targets, where the state variable represents the dynamic mode of the smart target and evolves through different transition matrices upon different actions.
We aim to maximize the discounted cumulative reward over an infinite time horizon, where the transition matrices for different actions are not known a priori.
Similar to the previous work mentioned above, we utilize the index policies and establish a priority-type scheduling framework, where the indices for each target are approximated through the difference between the Q values in active (tracked) and passive (untracked) actions.
The RMAB technique yields advantages in decomposing the large state space of the problem through the indices.
Instead of directly learning the Q values of the entire RMAB problem, we learn the Q values of each bandit process which can be utilized to approximate the indices accordingly.
To accelerate the convergence rate of the learning process, we adapt the Sarsa algorithm forward with sequential time episodes.
Based on the conventional Q learning steps, at the end of each episode, we adjust the Q values through a backward update by utilizing the state-action-reward-state tuples memory.
This framework iteratively updates the indices for each target and learns the dynamics of the smart target separately without memory sharing.
Unlike conventional techniques, it achieves decomposed learning and controlling across heterogeneous smart targets and proposes an expandable policy without incurring an excessively large amount of computational cost.
We refer to this policy as the index policy based on the Sarsa and Q-learning (ISQ).

Numerical results demonstrate that the ISQ policy outperforms all the tested state-of-the-art benchmarks, including the AB learning policy in \cite{avrachenkov2022whittle} and WIQL in \cite{biswas2021learn}, and that achieves the performance closest to the Whittle index policy with assumed full knowledge.
Our main contributions are summarized as follows.
\begin{enumerate}
\item
We formulate the multi-smart-target tracking problem in an APRN as an RMAB process with unknown transition matrices.
The dynamics of the smart target can transition into various candidate modes, e.g. constant velocity (CV) mode, constant acceleration (CA) mode, and constant turn (CT) mode. We set the discrete dynamic modes as the discrete states of each smart target.
We tackle the problem through a novel scheme with incorporated reinforcement learning and the index technique.
\item
To accelerate the low convergence rate, we embed the Sarsa algorithm in a Q-learning process, which imposes forward and backward steps of updating the estimated Q values in each time episode by utilizing the state-action-reward-state tuples memory.
With learned Q values, we are able to compute the state-dependent \emph{indices}.
We propose an \emph{index policy}, which is an online policy and is referred to as ISQ, for scheduling the beam resources of the APRN system.
\item
We numerically validate the Whittle indexability of the formulated RMAB problem through the \emph{strong indexability} discussed in \cite{nakhleh2021neurwin}.
Through extensive numerical results, we demonstrate the effectiveness of the ISQ policy in various scenarios, including those with classical circulant dynamics, homogeneous smart targets, and heterogeneous smart targets.
These tests collectively underscore the superior performance of the ISQ policy.
\end{enumerate}

The remainder of this paper is organized as follows.
In Section \ref{sec:model}, target dynamic models are defined and the corresponding MDP model is established.
The target tracking problem is formulated based on the RMAB model for non-myopic optimization.
In Section \ref{sec:solution}, we propose a priority-type index policy.
Following that, we propose the scheme incorporating the Sarsa and Q-learning techniques for our problem, leading to the ISQ policy.
In Section \ref{sec:simulation}, we conduct the numerical simulations that demonstrate the superiority of the ISQ policy through comparisons with benchmarks.
Section \ref{sec:conclusion} concludes this paper.

\section{Problem formulation}\label{sec:model}

\subsection{MDP-based modeling for smart targets}\label{subsec:IIA}
We consider an APRN, which consists of $K$ active radars and $N-K, (N >K)$ passive radars, to track $N$ smart targets.
Each radar can track only one target at each time.
Depending on whether being tracked or not by an active radar~\cite{savage2009sensor}, smart targets can change their dynamics into a set of candidate modes, such as CV mode, CA mode, and CT mode.
When being tracked by an active radar, a smart target will turn into a maneuvering mode, e.g., CA or CT, as a countermeasure against being monitored by the APRN.
Conversely, if the radiation from the active radar diminishes or the target is tracked by only passive radars, the target will gradually switch from the maneuvering mode to the CV mode.

The APRN has the ability to allocate active radars and passive radars to track smart targets.
In this setup, an active radar yields a higher reward, e.g., higher accuracy, compared to passive radars, while this advantage comes with a higher probability of missed detection, caused by evasive maneuvering.
Consequently, it becomes crucial to strike a balance between the tracking performance and the probability of missed detection.


We introduce the MDP model \cite{shi2021data} as a 6-tuple notation $(N, K, \mathcal{S}, \mathcal{A}, P^n, R)$ to model each target $n=1,2,\ldots,N$. In particular,
\begin{itemize}
\item Define $\mathcal{S}=\{0,1,\ldots, S-1\}$ as the set of states, where $S\in\mathbb{N}_+$, and
the target becomes more maneuverable as the state switches from 0 to $S-1$.
\item Let $\mathcal{A}$ denote a set of actions.
\item Define $P^n:\mathcal{S}\times\mathcal{A}\times\mathcal{S}\mapsto[0,1]$ as the transition kernel of the MDP associated with target $n = 1,2,\ldots,N$.
Recall that $P^n$ is not known a priori.
\item Define $R:\mathcal{S}\times\mathcal{A}\mapsto\mathbb{R}$ as the state-action-dependent reward function, where $\mathbb{R}$ is the set of real numbers.
\end{itemize}

\begin{figure}[!t]
\centering
    \begin{minipage}[t]{0.8\linewidth}
        \centering
        \includegraphics[width=\textwidth]{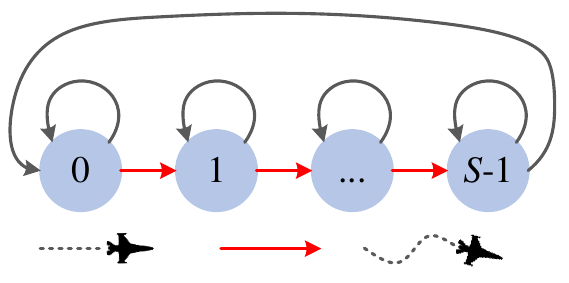}
        \centerline{(a)}
    \end{minipage}%

    \begin{minipage}[t]{0.8\linewidth}
        \centering
        \includegraphics[width=\textwidth]{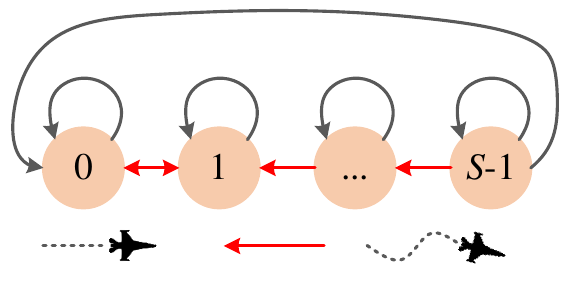}
        \centerline{(b)}
    \end{minipage}
    \caption{MDPs of smart targets under actions. (a) The tendency for maneuvering under action $a=1$. (b) The tendency for non-maneuvering under action $a=0$.}
    \label{fig_mdp}
\end{figure}

Denote the state of target $n$ at time $t$ as $X^n_t$, which is a random variable taking values in $\mathcal{S}$.
The action set $\mathcal{A}$ for each target $n$ includes two actions: being tracked by an active and a passive radar, represented by  $a^n_t=1$ or $0$, respectively.
The transition probability $P^n(X^n_{t+1}|X^n_t,a^n_t)$ of target $n$ represents the probability from the state $X^n_t$ to $X^n_{t+1}$ at time $t+1$ under action $a^n_t$, and
$R(X^n_t,a^n_t)$ denotes the reward obtained under action $a^n_t$ when target $n$ is in state $X^n_t$.
Fig.~\ref{fig_mdp} illustrates the state transitions of target $n$ under different actions, where, recall that, from state $0$ to $S-1$, the probability for the target to change into a maneuvering mode increases.
That is, if the target is tracked by an active radar ($a^n_t=1$), then it will either stay in its current state or transition to a state with a higher probability for maneuvering; otherwise, the target will progressively switch to state $0$ - the state with the least probability for being maneuvering or in a non-maneuvering mode.

\subsection{Problem formulation}\label{subsec:IIB}
Recall that there are $K<N$ active radars in the APRN system, for which we impose the following constraint
\begin{equation}\label{eq1}
\sum_{n=1}^N a^n_t = K,~t=1,2,\ldots,\infty.
\end{equation}

The goal is to find an optimal policy $\pi^*$, comprised of the actions $a^n_t$ for all $n=1,2,\ldots,N$ and $t=1,2,\ldots$, that maximizes the expected discounted cumulative reward, as given by
\begin{equation}\label{eq2}
\begin{aligned}
\max_{\pi} &~\mathrm{E}^{\pi}\left[ \sum_{t=0}^{\infty}\sum_{n=1}^N \beta^t R(X^n_t,a^n_t) \right], \\
& \text{s.t.}~\eqref{eq1},
\end{aligned}
\end{equation}
where $\beta\in(0,1)$ is the discounting factor, and $\mathrm{E}^{\pi}[\cdot]$ is the expectation operation under the given policy $\pi$.
Recall that the behavior of smart targets tends to be elusive and convoluted, making the state transition kernel ${P}^n$ unpredictable and hardly known a priori.
A novel radar scheduling scheme, which is equipped with both efficient and adaptive learning attributes, becomes imperative.

\section{Sarsa and Q-learning based index policy}\label{sec:solution}
The problem \eqref{eq2} is a standard RMAB problem, where the size of the state space increases exponentially in the number of smart targets.
A standard RMAB with known transition matrices can be in general addressed by the well-known \emph{index policy}.
However, in our case, the transition kernel is unknown a priori.
We incorporate the Sarsa and Q-learning techniques with the RMAB technique, enhancing the overall performance by appropriately learning the unknown information.
In this manner, we propose the ISQ policy that tracks heterogeneous smart targets in an APRN by appropriately scheduling the radar beams.

\subsection{Index policy}\label{subsec:IIIA}
\begin{figure}[!t]
\centering
\includegraphics[width=0.8\linewidth]{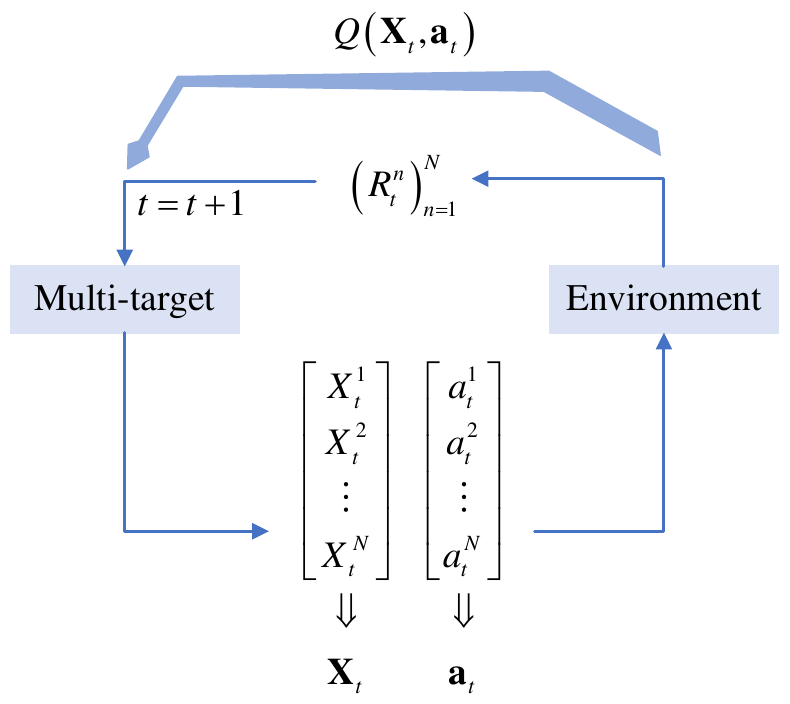}
\caption{The framework of Reinforcement Learning.}
\label{figframeQ}
\end{figure}

Conventional reinforcement learning algorithms model the multi-target scheduling problem as an MDP, where the joint state $\mathbf{X}_t\triangleq (X^n_t)^N_{n=1}$ consists of individual states $X^n_t$ of targets $n=1,2,\ldots,N$, and the action vector $\mathbf{a}_t\triangleq (a^n_t)^N_{n=1}$ is decided by the action variable $a^n_t$ for each target $n$ at time $t$.
Subject to constraint \eqref{eq1}, define the feasible action set $\bm{\mathcal{A}}\triangleq \mathcal{A}_1 \times \mathcal{A}_2 \ldots\times \mathcal{A}_N$, conditioned that only $K$ elements in $\mathbf{a}_t$ can be set to 1.
The Q value is a function $Q(\mathbf{X}_t,\mathbf{a}_t)$ of the state-action pairs.
The framework of reinforcement learning algorithms is shown in Fig.~\ref{figframeQ}.

With the constraint \eqref{eq1}, Problem \eqref{eq2} can be transformed to
\begin{equation}\label{eqRL}
\begin{aligned}
    \max_{\pi} &~\mathrm{E}\left[ \sum_{t=0}^{\infty}\sum_{n=1}^N \beta^t R(X^n_t,a^n_t) \right], \\
    & \text{s.t.}~ \mathbf{a}_t\in \bm{\mathcal{A}}.
\end{aligned}
\end{equation}

Given $\bm{\mathcal{A}}$, reinforcement learning algorithms can iteratively learn the Q values through interactions with the MTT environment. Nonetheless, in general, the sizes of the state and action spaces reversely affect the convergence speed of the learning process.
In our case, the sizes of the state and action spaces are $S^N$ and $\binom{N}{K}$, respectively, increasing fast as $N$ becomes large. causing a low convergence rate of $Q(\mathbf{X}_t,\mathbf{a}_t)$ through conventional reinforcement learning techniques.

For a conventional RMAB problem with full knowledge of the transition matrices, the Whittle index policy is proved to be asymptotically optimal \cite{whittle1988restless,weber1990index} in certain cases if the RMAB is indexable, which has been considered as a vital property for analyzing RMAB problems.
The details of the Whittle index policy and indexability are provided in Appendix \ref{AppWI} and Appendix \ref{App1}, respectively.
When the transition matrices are unknown, \cite{fu2019towards,avrachenkov2022whittle} approximated Whittle indices using Q-learning in a coupling iteration manner, where the Q values of each target consist of the estimated Whittle index.
However, such a learning scheme may reversely impact the overall performance of the employed scheduling policies.

In this paper, we develop an index policy to define the distinctive index $\lambda^{n*}(X^n_t)$ by the difference between Q values $Q^n(X^n_t,a^n_t)$ corresponding to action 1 and action 0, following \cite{biswas2021learn}. The index on state $X^n_t$ for target $n$ is given by
\begin{equation}\label{eqlambda}
\lambda^{n*}(X^n_t)\triangleq Q^{n*}(X^n_t,1)-Q^{n*}(X^n_t,0),
\end{equation}
where $Q^{n*}(X^n_t,a^n_t)$ denotes the optimal Q value to compute the discounted rewards for target $n$.

\begin{figure}[!t]
\centering
\includegraphics[width=0.8\linewidth]{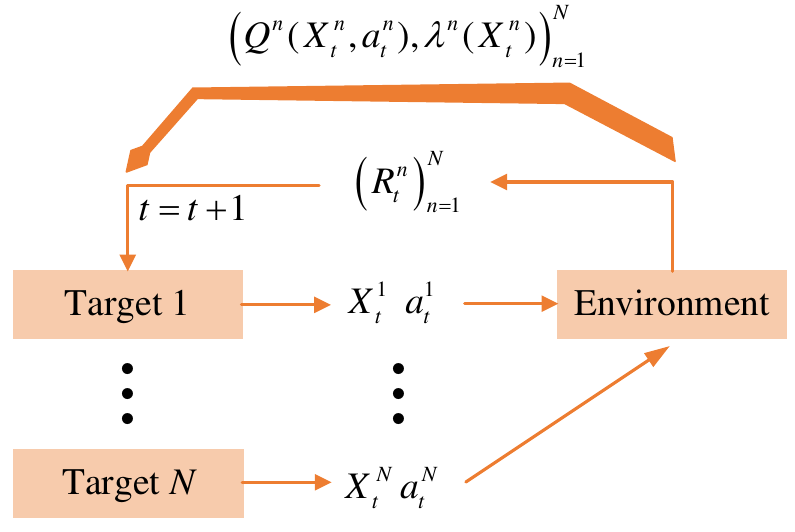}
\caption{The framework of the index policy based on reinforcement learning.}
\label{figframeWI}
\end{figure}

Such indices quantify the priorities of the targets when they are in various states, and, along with the constraint \eqref{eq1}, the index policy will track the $K$ smart targets according to their index values at each time $t$.
Problem \eqref{eq2} can be transformed into a solution with the computational complexity linear in $N$.
We adopt the reinforcement learning algorithm for each target to accelerate the learning rates of the conventional off-the-shelf learning technique while mitigating the interference caused by simultaneous exploration and exploitation.
The reinforcement-learning-based index policy independently updates the Q values $Q^n(X, a)$ for each target during the learning process.
Meanwhile, the index $\lambda^n(X)$ is updated based on the computed Q values at each time slot, as illustrated in Fig.~\ref{figframeWI}.
In this context,
we conduct the learning process with $N$ independent MDPs, each of which corresponds to a target with $S$ and $2$ the sizes of its state and action spaces, respectively.

\subsection{Sarsa algorithm - forward learning}\label{subsec:IIIB}
Sarsa \cite{van2009theoretical,aslam2019optimal} and Q-Learning \cite{watkins1992q} are well-known RL algorithms for estimating optimal Q values concerning specific state-action pairs. The optimal Q values of target $n$ can be expressed through the following Bellman equation,
\begin{equation}\label{eq5}
\begin{aligned}
Q^{n*}(X^n_{t},a^n_t)
=& a^n_t\left[R(X^n_t,a^n_t)+\beta\bar{V}^{n*}(X^n_t,a^n_t)\right] \\
&+(1-a^n_t)\left[R(X^n_t,a^n_t)+\beta\bar{V}^{n*}(X^n_t,a^n_t)\right]\\
=& a^n_t R(X^n_t,a^n_t) + (1-a^n_t)R(X^n_t,a^n_t) \\
& + \beta\bar{V}^{n*}(X^n_t,a^n_t),
\end{aligned}
\end{equation}
where
\begin{equation}\label{eq51}
\begin{aligned}
\bar{V}^{n*}(X^n_t,a^n_t)
=\sum_{X^n_{t+1}\in\mathcal{S}} P^n(X^n_{t+1}|X^n_{t},a^n_t) V^{n*}(X^n_{t+1}),
\end{aligned}
\end{equation}
and
\begin{equation}\label{eq52}
\begin{aligned}
V^{n*}(X^n_{t+1})=\max_{b\in{\{0,1\}}}Q^{n*}(X^n_{t+1},b).
\end{aligned}
\end{equation}

Recall that the transition matrix $P^n$ of target $n$, $n=1,\ldots, N$ is unknown to APRN.
To address this issue, the Sarsa and Q-learning algorithms, based on the temporal-difference (TD) learning method \cite{bertsekas1996neuro}, are adopted to facilitate the acquisition of Q values that exhibit convergence to the optimal ones through a series of episodes.
While Q-learning selects the subsequent action by maximizing Q values in the subsequent state, Sarsa selects the subsequent actions by the $\epsilon$-decay method based on learned Q values at the current time slot \cite{wang2013backward}.
In \cite{sutton2018reinforcement}, Sarsa guarantees the convergence of the Q-learning process and highlights a faster convergence rate than conventional Q-learning, as demonstrated in experimental results in the cliff-walk problems.

Given the above-mentioned advantages of Sarsa, here, we adapt Sarsa to construct a \emph{forward learning} process for each smart target across each learning episode consisting of successive time slots, each of which triggers index updates.

More precisely, the Q values of target $n$ at time $t+1$ are estimated by the Sarsa algorithm \cite{al2020reinforcement} as follows.
Let
\begin{equation}\label{eq8}
\begin{aligned}
&\hat{Q}^{n}_{t+1}(X^n_{t},a^n_t) \\
=&(1-\alpha^n_t(L^n_t(X^n_{t},a^n_t)))\hat{Q}^{n}_t(X^n_{t},a^n_t) \\
&+\alpha^n_t(L^n_t(X^n_{t},a^n_t))\left[ R(X^n_t,a^n_t) + \beta\hat{Q}^{n}_{t}(X^n_{t+1},a^n_{t+1})\right],
\end{aligned}
\end{equation}
where action $a^n_{t+1}$ is chosen from state $X^n_{t+1}$ using the $\epsilon$-decay~\cite{sutton2018reinforcement}.
In our paper, the $\epsilon$-decay is triggered every time slot $t$ and is processed based on the updated indices $(\hat{\lambda}^n_{t-1}(X^n_{t}))_{n=1}^N$ in the previous time slot $t-1$.
The learning ratio $\alpha^n_t$ should satisfy the conditions $\sum_{t}^{\infty}\alpha^n_t=\infty$, $\sum_{t}^{\infty}(\alpha^n_t)^2<\infty$ to guarantee the learning convergence~\cite{avrachenkov2022whittle}.
Define
\begin{equation}\label{eq81}
\alpha^n_t(L^n_t(X^n_{t},a^n_t))=\frac{1}{L^n_t(X^n_{t},a^n_t)+1},
\end{equation}
where $L^n_t(X^n_{t},a^n_t)=\sum_{i=0}^t\mathbb{I}(X^n_{i}=X^n_{t},a^n_{i}=a^n_t)$, and $\mathbb{I}(\cdot)$ is the indicator function.

Based on the estimated Q values, we update the index $\hat{\lambda}^n_{t+1}(X^n_{t})$ at time slot $t+1$,
\begin{equation}\label{eq9}
\begin{aligned}
\hat{\lambda}^n_{t+1}(X^n_{t})
=\hat{Q}^{n}_{t+1}(X^n_{t},1)-\hat{Q}^{n}_{t+1}(X^n_{t},0).
\end{aligned}
\end{equation}

The Sarsa algorithm is used to learn the Q values forward in each episode, which consists of successive time slots, and the indices are estimated upon the most updated Q values on actions 1 and 0.
The estimated indices quantify the marginal rewards of tracking certain targets, in particular, constructing an expandable policy that prioritizes the $K$ targets according to the descending order of their time-variant indices.
Nonetheless, in this context, some targets may be inappropriately omitted due to poor index values at the beginning.
When consistently overlooking certain targets, the unchanged Q values may also negatively affect the convergence of the true indices.

We mitigate the unfairness through $\epsilon$-decay \cite{biswas2021learn}.
Along the idea of the $\epsilon$-decay, we introduce a $\epsilon_t$ factor to denote the probability of exploration at each time slot, and the factor will decrease as the learning process evolves over time.
With probability $\epsilon_t$, we select $K$ targets uniformly at random. With probability $1-\epsilon_t$, select top $K$ arms according to their index values $(\hat{\lambda}^n_{t})^N_{n=1}$.
This method ensures a heightened emphasis on exploration during the initial learning stage and gradually transitions towards a pronounced emphasis on exploitation as the learning process matures. In particular, the $\epsilon_t$ factor is given by,
\begin{equation}\label{eqep}
\epsilon_t=e/(e+t),
\end{equation}
where $e$ is a constant.


\subsection{Q-learning algorithm - backward learning} \label{subsec:IIIC}
In pursuit of enhancing the aforementioned learning of Q values, we adopt the Q-learning technique to facilitate Q values and the indices backward at the end of each episode.
This process is grounded in the utilization of memory information $\mathcal{H}^n_j$, which represents the history of state-action-reward-state tuples of target $n$ in the episode $j$, $j=1,\ldots, J$.
For each $j$, $\mathcal{H}^n_j$ includes $T$ tuples $H^n_{j,t}=\{X^n_t,a^n_t, R(X^n_t,a^n_t),X^n_{t+1}\}$, $t=0,\ldots,T-1$, where $T$ is the number of time slots included in each episode and is referred to as the \emph{length} of the episode.

The backward Q-learning is triggered at the end time slot $T-1$ of episode $j$ and is handled with $\mathcal{H}^n_j$.
In this sub-section, we consider the Q values and indices for time slot $t=T-1$ and, for simplicity of notation, omit the subscript $t$ for $\hat{Q}^{n}_t$ and $\hat{\lambda}^n_t$.
The Q values are updated as follows
\begin{equation}\label{eq10}
\begin{aligned}
\hat{Q}^{n}(X^n_t,a^n_t)
\leftarrow&(1-\bar{\alpha})\hat{Q}^{n}(X^n_t,a^n_t) \\
&+ \bar{\alpha}\left[R(X^n_t,a^n_t)
+ \beta \max_{b\in\mathcal{A}}\hat{Q}^{n}(X^n_{t+1},b) \right],
\end{aligned}
\end{equation}
where $\bar{\alpha}$ is a constant learning rate.
In the backward learning process, we update the index based on $H^n_{j,t}$,
\begin{equation}\label{eq11}
\begin{aligned}
\hat{\lambda}^n(X^n_{t})
= \hat{Q}^{n}(X^n_{t},1)-\hat{Q}^{n}(X^n_{t},0).
\end{aligned}
\end{equation}

Through $T$ backward iterations within an episode $j$, we obtain the newly learned Q values $\hat{Q}^{n}$ and the updated indices $\hat{\lambda}^n$.
Assign the values of $\hat{Q}^{n}$ and $\hat{\lambda}^n$ to $\hat{Q}^{n}_{t_0}$ and $\hat{\lambda}^n_{t_0}$, respectively, for all $n=1,2,\ldots,N$, where $t_0=0$ is the starting time slot for next episode $j+1$.

\subsection{ISQ scheduling policy}\label{subsec:IIID}
\begin{figure}[!t]
\centering
\includegraphics[width=\linewidth]{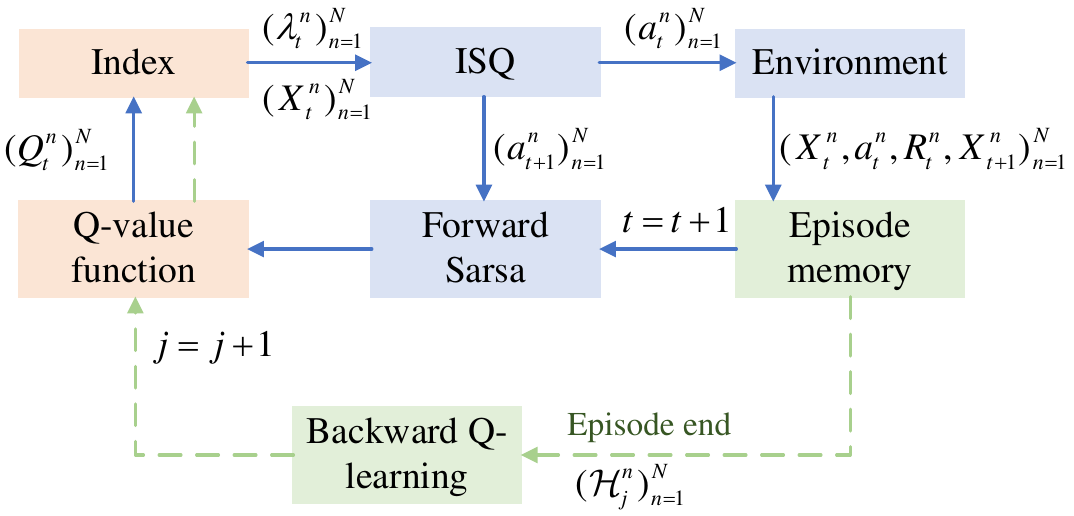}
\caption{The proposed framework of radars scheduling in the APRN.}
\label{figframe}
\end{figure}

As described in the above sections, we adapt the index policy with incorporated Sarsa and Q-learning techniques and refer to this entire scheme as the index policy based on Sarsa and Q-learning (ISQ) policy.
The ISQ policy achieves effective estimation of Q values through the forward and backward learning processes, enhancing both the rate of convergence and the precision of the estimated Q values.
It follows with effectively learned indices that approximate the marginal rewards of the state-target pairs.
We depict the framework of the ISQ policy in Fig.~\ref{figframe}.
In the forward learning process, after estimating the Q values and the indices through Sarsa, the $K$ smart targets with the $K$ updated largest indices will be selected for active tracking with probability $1-\epsilon_t$.
This forward process is illustrated using solid blue lines.
While in the backward learning process, we update the Q values and the indices based on the memory information within the current episode. The backward process is depicted by dotted green lines in Fig.~\ref{figframe}.
Also, the pseudo-code of the ISQ policy is provided in Algorithm \ref{algorithm:Algorithm 1}.
The computational complexity for the entire process for each episode is $O(TN)$, where $T$ is the length of each episode and $N$ is the number of targets.

\begin{algorithm}[!t]
\DontPrintSemicolon
\SetAlgoLined
\caption{The scheduling scheme based on the ISQ policy}
\label{algorithm:Algorithm 1}
\tcp*[l]{\textbf{Initialization}}
The problem model parameters, i.e. $N$, $K$, $J$, $T$, $\beta$, and the target parameters, i.e. $\mathcal{S}$, $\mathcal{A}$, $\mathcal{P}$, $R$. \\
$\hat{Q}^{n}_0(X,a)\gets R(X,a)$, $\hat{\lambda}^n_0\gets0$, $L^n_0\gets0$ for each state $x\in\mathcal{S}$, each action $a\in\mathcal{A}$, and each arm $n=1,\ldots,N$.\\
\tcp*[l]{\textbf{Main loop}}
\For{$j \gets 1$ \KwTo $J$}{
Random initial states $X^n_0\in\mathcal{S}$ for each arm;\;
With probability $\epsilon_t=e/(e+t)$, select $K$ targets uniformly at random, where $e$ is a constant parameters and $t=0$. Otherwise, select the $K$ largest targets with the $K$ largest index $\hat{\lambda}^n_{0}$;\;
\tcp*[l]{\textbf{Forward learning}}
\For{$t \gets 0$ \KwTo $T-1$}{
\For{$n \gets 1$ \KwTo $N$}{
Execute actions on targets $a^n_t$, observe the reward $R(X^n_t,a^n_t)$, and next state $X^n_{t+1}$;\;
Update $L^n_t(X^n_{t},a^n_t)$;\;
}
Choose actions $(a^n_{t+1})^N_{n=1}$ using the $\epsilon$-decay policy;\;
\For{$n \gets 1$ \KwTo $N$}{
Record the states and actions in $\mathcal{H}^n_{j,t}$;\;
Update $\hat{Q}^{n}_{t+1}$ and $\hat{\lambda}^n_{t+1}$ in \eqref{eq8} and \eqref{eq9}, respectively;\;
}
Calculate the reward in time $t$.
}
\tcp*[l]{\textbf{Backward learning}}
\For{$t \gets T-1$ \KwTo $0$}{
\For{$n \gets 1$ \KwTo $N$}{
The backward Q-learning for updating $\hat{Q}^{n}$ and $\hat{\lambda}^n$ in \eqref{eq10} and \eqref{eq11} with $\mathcal{H}^n_{j,t}$, respectively.\;
}
}
$\hat{Q}^{n}_0\gets \hat{Q}^{n}$, $\hat{\lambda}^n_0\gets\hat{\lambda}^n$, $L^n_0\gets L^n_{T-1}$.\\
}
\end{algorithm}

For the case with one MDP/target, following the convergence proofs of learned Q values of the Sarsa and Q-learning algorithms in \cite{singh2000convergence,bertsekas1996neuro,van2009theoretical},
the updated Q value converges with probability 1 to the optimal Q value, if the learning rate $\alpha_t$ satisfies the conditions: $\sum_{t}^{\infty}\alpha_t=\infty$, $\sum_{t}^{\infty}(\alpha_t)^2<\infty$.
In this paper, we define the learning rate $\alpha^n_t$ for each target $n$ in \eqref{eq81}, subject to the conditions: $\sum_{t}^{\infty}\alpha^n_t=\infty$, $\sum_{t}^{\infty}(\alpha^n_t)^2<\infty$.
Certainly, the $\epsilon$-decay policy guarantees that all state-action pairs are visited frequently in the forward learning process.

For the general case with multiple targets, when the Q values converge to the optimal value $Q^{n*}(X^n_t,a^n_t)$ for each target, the ISQ policy based on $\lambda^{n*}(X^n_t)$ computes the joint action $\mathbf{a}_t=(a^n_t)^N_{n=1}$ at each time subject to $\sum^N_{n=1}a^n_t = K$.
Nonetheless, for the entire process with mutually affected learning and controlling, the convergence of the learned $\hat{Q}^n_t(X^n_t,a^n_t)$ and, more importantly, the deviation between $\hat{Q}^n_t(X^n_t,a^n_t)$ and the ideal $Q^{n*}(X^n_t, a^n_t)$ for each target $n$ remain open questions.
Also, in general, even if $\hat{Q}^n_t(X^n_t, a^n_t)$ converges to $Q^{n*}(X^n_t, a^n_t)$ as $t\rightarrow \infty$, it is unclear whether or not the corresponding $Q^{n*}(X^n_t, a^n_t)$ converges to the optimal Q value, $Q^{*}(\mathbf{X}_t, \mathbf{a}_t)$, of Problem~\eqref{eqRL}.

Eventually, given the learned Q value $Q^{n}_t(X^n_t,a^n_t)$ for $X^n_t\in\mathcal{S}$, under ISQ, the selected joint action $\mathbf{a}_t$ maximizes $\max_{\mathbf{a}_t\in\{0,1\}^N: \sum^N_{n=1}a^n_t = K}\bar{Q}_t(\mathbf{X}_t,\mathbf{a}_t)$ where $\bar{Q}_t(\mathbf{X}_t,\mathbf{a}_t)\triangleq \sum^N_{n=1}Q^{n}_t(X^n_t, a^n_t)$.
It takes consideration of current and future gains when taking certain actions across all the targets and prioritizes those targets with the highest marginal rewards - the differences between their estimated Q values with active and passive actions.
We argue that the ISQ policy approximates the optimality of the overall MTT problem, and, in Section~\ref{sec:simulation}, we numerically demonstrate the effectiveness of ISQ through extensive simulations.

\section{Simulation and Analysis}\label{sec:simulation}
We numerically demonstrate the effectiveness of the proposed algorithm through three simulation cases with 20 trials, where the first case is with the circulant dynamics from \cite{avrachenkov2022whittle,fu2019towards}, and the other two are based on the smart target tracking problem with homogeneous and heterogeneous cases, respectively.
We compare the proposed ISQ policy with the following four benchmarks.
(1) Whittle index (WI) policy, which assumes full knowledge of transition matrices, is considered as a performance upper bound.
The strong indexability of the corresponding RMAB problem is validated numerically - the problem is indexable. The details on how to validate Whittle indexibility and strong indexibility are provided in Appendices~\ref{AppWI} and \ref{App1}, respectively.
(2) Whittle index-based Q-learning (WIQL) policy was proposed in~\cite{biswas2021learn}, where $\epsilon_t=N / (N + t)$.
(3) The AB policy was studied in~\cite{avrachenkov2022whittle}, for which we set $\epsilon=0.01$, $C=1/5$, and $C'=1/3$.
In the last two cases based on smart target tracking, finding the best hyper-parameters for the AB policy is not straightforward, and, in the third case with heterogeneous smart targets, the AB policy is not applicable.
As a result, for the last two cases, we exclude the AB policy for comparison.
(4) Greedy policy~\cite{biswas2021learn} assumes the expected instantaneous rewards for each arm (target) under given states and actions and, after observing the states of all arms at each decision epoch, greedily selects the targets with the highest differences between the expected instantaneous rewards of taking 1 and 0 actions.

\subsection{Circulant dynamics}\label{subsec:IVA}
Consider a system that comprises a set of $N$ homogeneous arms.
At each time slot, $K$ arms are selected to maximize the time-averaged reward.
The state space is defined as $\mathcal{S}=\{0,1,2,3\}$.
The state of arm $n$ evolves according to transition probability matrices under action $a^n_t\in\mathcal{A}=\{0,1\}$, which are given by
\begin{equation}\label{eqA1}
\begin{aligned}
P^n(X^n_{t+1}\!\in\!\mathcal{S}|X^n_t\!\in\!\mathcal{S},a^n_t\!=\!0)\!=\! \begin{bmatrix}
0.5 & 0 & 0   & 0.5 \\
0.5 & 0.5   & 0   & 0 \\
0 & 0.5 & 0.5   & 0 \\
0   & 0 & 0.5 & 0.5
\end{bmatrix},
\end{aligned}
\end{equation}
\begin{equation}\label{eqA2}
\begin{aligned}
P^n(X^n_{t+1}\!\in\!\mathcal{S}|X^n_t\!\in\!\mathcal{S},a^n_t\!=\!1)\!=\!\begin{bmatrix}
0.5 & 0.5 & 0   & 0 \\
0 & 0.5   & 0.5   & 0 \\
0 & 0 & 0.5   & 0.5 \\
0.5   & 0 & 0 & 0.5
\end{bmatrix},
\end{aligned}
\end{equation}
where the $(q,v)$th entry of $P^n$ denotes the transition probability of arm $n$ from state $q\in\{0,1,2,3\}$ to state $v\in\{0,1,2,3\}$ at time $t$ under action $a_t^n$.

The expected instantaneous rewards with given states and actions are
$R(X^n_t=0,a^n_t\in\mathcal{A})=-1$, $R(X^n_t=1,a^n_t\in\mathcal{A})=R(X^n_t=2,a^n_t\in\mathcal{A})=0$, and $R(X^n_t=3,a^n_t\in\mathcal{A})=1$.

The exact values of the Whittle indices are considered the same as those in~\cite{fu2019towards}: $\lambda^*(0)=-0.5$, $\lambda^*(1)=0.5$, $\lambda^*(2)=1$, and $\lambda^*(3)=-1$.
In this case, the Whittle index policy gives priority to state $2$.

\begin{figure}[!t]
\centering
    \begin{minipage}[t]{0.45\textwidth}
        \centering
        \includegraphics[width=\textwidth]{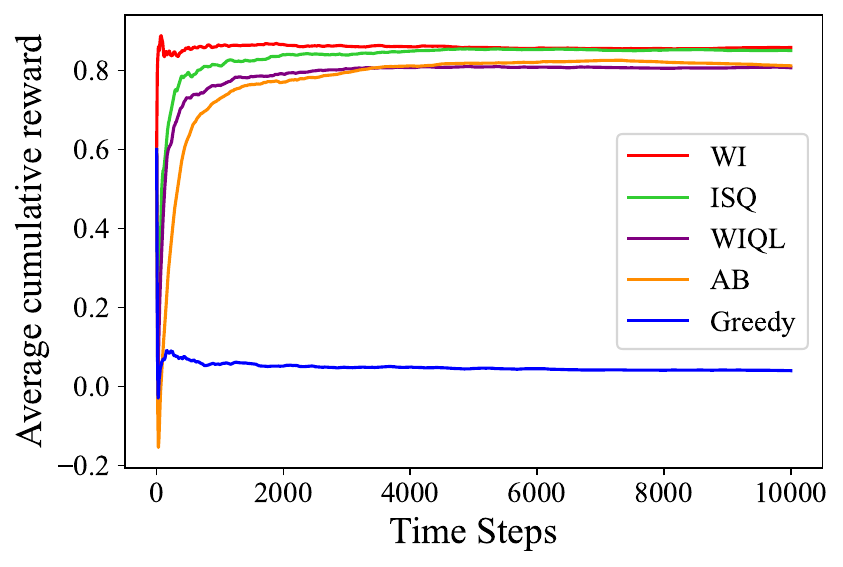}
        \centerline{(a)}
    \end{minipage}%

    \begin{minipage}[t]{0.45\textwidth}
        \centering
        \includegraphics[width=\textwidth]{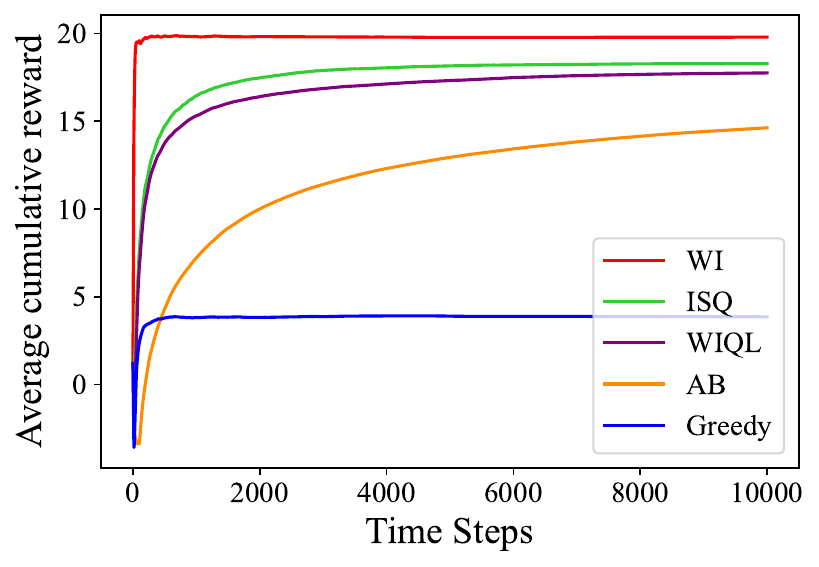}
        \centerline{(b)}
    \end{minipage}
    \caption{Average cumulative reward for the circulant dynamics problem. (a) $N=5$, $K=1$. (b) $N=100$, $K=20$.}
    \label{fig1}
\end{figure}

We set the exploration ratio as $\epsilon_t = N/(N + t)/2$ in the $\epsilon$-decay policy of ISQ.
We run ISQ and the four benchmarks.
The results of cumulative time-average reward when $N=5$, $K=1$ and $N=100$, $K=20$ are shown in Fig.~\ref{fig1}, where the Whittle Index policy is abbreviated as WI.
In the case with $N=5$, $K=1$, WIQL and AB policies fail to converge to the performance of the WI policy.
The Greedy policy has the worst performance, scoring about 0, due to its inherent myopic limitation.
The ISQ policy can accelerate the convergence of learning during the initial learning stage and ultimately converge to the performance of the WI policy.
With the increased scales $N=100$, $K=20$, certain arms encounter limited learning opportunities without memory sharing.
Consequently, they receive reduced feedback information within the same time horizon as the $N=5$, $K=1$ case.
Despite a reduction in performance~($7\%$), ISQ obtains the nearest performance to the WI policy and again outperforms the other three benchmark policies.
In summation, the ISQ policy demonstrates its fast convergence and enhanced performance in the circulant dynamic problem.

\subsection{Homogeneous smart targets tracking}\label{subsec:IVB}

In this case, we consider an APRN comprising $K$ active radars and $N-K$ passive radars to track $N$ homogeneous smart targets.
To highlight the long-run effects of the employed actions, we take $\beta=0.999$ in this subsection.
The state space is defined as $\mathcal{S}=\{0,1,2,3\}$ and represents $\{\mathrm{CV}, \mathrm{CA}, \mathrm{CT}, \mathrm{NT}\}$, respectively, where ``NT" means the target is not detected.
An action $a^n_t\in\mathcal{A}=\{0,1\}$ is taken for target $n$ at time slot $t$. Here, $a^n_t = 0$ signifies passive tracking, while $a^n_t = 1$ corresponds to active tracking.
The transition matrices of target $n$ are
\begin{equation}\label{eqtarget1}
\begin{aligned}
P^n(X^n_{t+1}\!\in\!\mathcal{S}|X^n_t\!\in\!\mathcal{S},a^n_t\!=\!0) \!= \! \begin{bmatrix}
0.8 & 0.2 & 0   & 0 \\
0.3 & 0.7 & 0   & 0 \\
0   & 0.3 & 0.7 & 0 \\
0.4 & 0   & 0   & 0.6
\end{bmatrix}, \\
\end{aligned}
\end{equation}
\begin{equation}\label{eqtarget2}
\begin{aligned}
P^n(X^n_{t+1}\!\in\!\mathcal{S}|X^n_t\!\in\!\mathcal{S},a^n_t\!=\!1)\! =\! \begin{bmatrix}
0.3 & 0.7 & 0   & 0 \\
0   & 0.3 & 0.7 & 0 \\
0   & 0   & 0.3 & 0.7 \\
0.3 & 0   & 0   & 0.7
\end{bmatrix}.
\end{aligned}
\end{equation}

In general, given a target state $X$, the tracking reward under $a=0$ is smaller than that under $a=1$.
However, the active tracking action may potentially induce maneuvering behaviors of smart targets, even missed detection of the APRN.
The absence of radiation from passive radars precludes smart target states from state ``NT".
In light of these considerations, we consider the reward decreases as the state transitions from 0 to 3; that is,
\begin{equation}\label{eq13}
R(X^n_t\in\mathcal{S},a^n_t\in\mathcal{A})= \begin{bmatrix}
0.5 & 0.3 & 0.1 & 0 \\
2   & 1.5 & 1   & -1
\end{bmatrix}',
\end{equation}
where $R(X^n_t,0)<R(X^n_t,1)$, $X^n_t\in\{0,1,2\}$, represents that the tracking reward of passive tracking is smaller than that of active tracking, excluding state 3.
When $X^n_t = 3$ and $a_t^n = 1$, the reward is assumed to be negative, which serves as a punishment for the missed detection of smart targets.
$[\cdot]'$ denotes the transpose operator of a matrix.

We compute the exact Whittle index using full knowledge of the transition matrices, which remain unknown for the other tested policies.
The Whittle indices are $\lambda^*(0)=1.3060$, $\lambda^*(1)=0.4129$, $\lambda^*(2)=1.0237$ and $\lambda^*(3)=-1.4711$.
These Whittle indices show that the smart target with state 0 is more likely to be tracked by active radars, attributed to its non-maneuvering behavior.
When a target transitions to state 3 (``NT''), its Whittle index becomes negative due to the risk of maneuvering and being non-detected under active tracking.
For state 3 (``NT'') and action 1 (being actively tracked), the smart target exhibits a state transition probability of 0.7 to stay in state 3 (``NT'').
The negative index $\lambda^*(3)=-1.4711$ indicates that the APRN refrains from active tracking.
Recall that active tracking implies higher tracking rewards but, in general, is more risky for triggering maneuver behaviors, resulting in performance degradation in future time slots.
While imperceptible passive tracking may gain fewer tracking rewards at the current time, it avoids maneuvering and may ensure more stable gain in the long run.
The Whittle index policy takes consideration of the marginal rewards for different actions and targets in a long-run manner, which in general demonstrates advantages against myopic policies.

\begin{figure}[!t]
\centering
\includegraphics[width=0.45\textwidth]{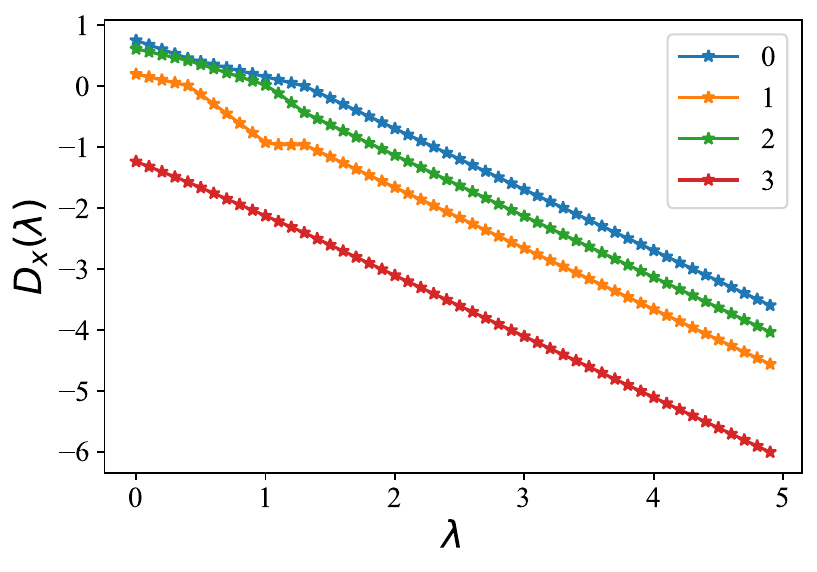}
\caption{Difference $D_X(\lambda)$ in diverse $X$.}
\label{D(X)}
\end{figure}

We employ the value iteration method with the known state transitions, aiming to compute the Q values of the optimal policy in diverse $\lambda$ and calculate the difference $D_X(\lambda)$ of states $X$ in $\lambda$. The results are shown in Fig.~\ref{D(X)}, where $D_X(\lambda)$ of all four states are strictly decreasing in $\lambda$.
Meanwhile, we numerically verified that the strong indexability condition in Definition \ref{d2} of Appendix \ref{App1} is achieved - the tested smart target problem is indexable for using the WI policy.

\begin{figure}[!t]
\centering
    \begin{minipage}[t]{0.45\textwidth}
        \centering
        \includegraphics[width=\textwidth]{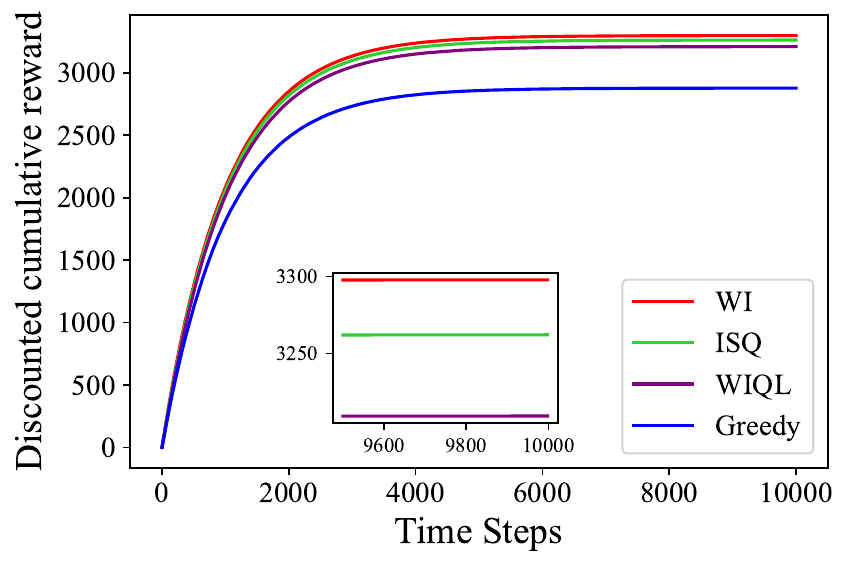}
        \centerline{(a)}
    \end{minipage}%

    \begin{minipage}[t]{0.45\textwidth}
        \centering
        \includegraphics[width=\textwidth]{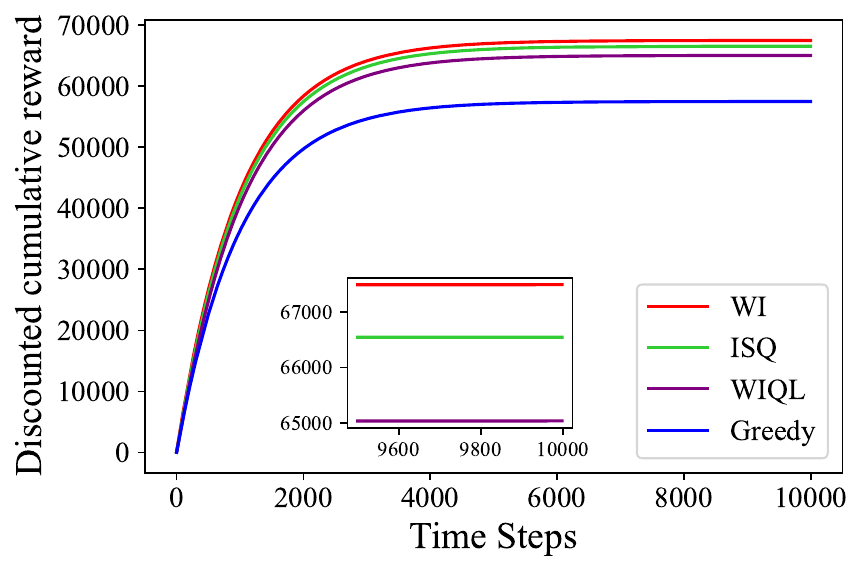}
        \centerline{(b)}
    \end{minipage}
    \caption{Discounted cumulative reward for homogeneous smart targets. (a) $N=5$, $K=1$. (b) $N=100$, $K=20$.}
    \label{fig2}
\end{figure}

For  $N=5$, $K=1$ and $N=100$, $K=20$, the discounted cumulative rewards of ISQ and the three benchmark policies are shown in Fig.~\ref{fig2}.
Leveraging an accelerated learning rate during the early learning stage, ISQ again obtains a higher discounted cumulative reward, which is the closest to that of WI.
In particular, when $N=100$, $K=20$, the performance difference between ISQ and WIQL increases significantly, up to 1500.
In comparison to WIQL, the ISQ policy achieves improvements of approximately 3.16\% and 2.31\% higher rewards, respectively. However, the Greedy policy yields inferior performance when compared to both WIQL and ISQ.

\subsection{Heterogeneous smart targets tracking}\label{subsec:IVC}
Similar to the second case in Section~\ref{subsec:IVB}, we consider the same settings except for distinct state transitions across all the smart targets - the targets are heterogeneous.
We uniformly randomly generate the entries of $P^n$ in \eqref{eqtarget1} and \eqref{eqtarget2}.
The generation process adheres to a specific criterion, where the smart target has a higher probability to maneuver when being actively tracked; meanwhile, it has a higher probability to gradually transition towards the CV dynamic model under the passive tracking.
The reward function remains consistent with \eqref{eq13}.
Regarding the ISQ policy, we set $e=5$ for the parameter $e$ in \eqref{eqep}.

\begin{figure}[!t]
\centering
    \begin{minipage}[t]{0.45\textwidth}
        \centering
        \includegraphics[width=\textwidth]{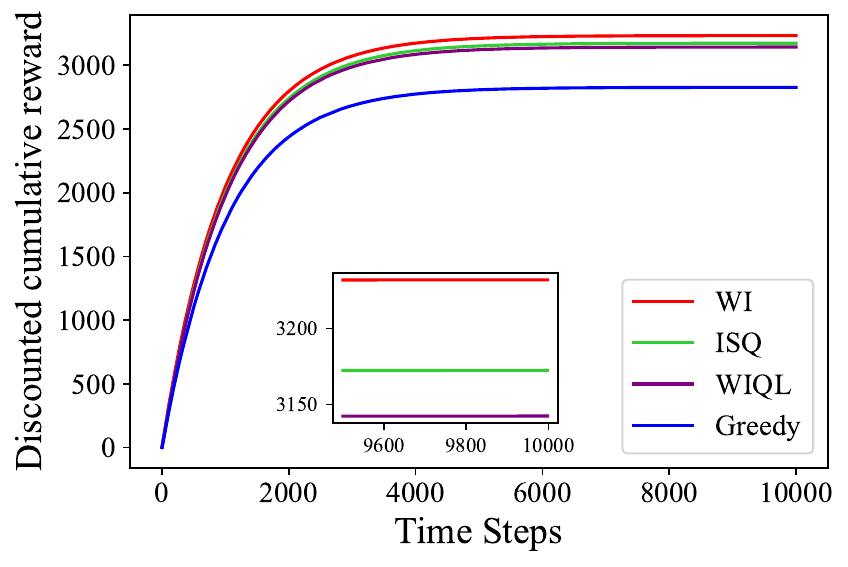}
        \centerline{(a)}
    \end{minipage}%

    \begin{minipage}[t]{0.45\textwidth}
        \centering
        \includegraphics[width=\textwidth]{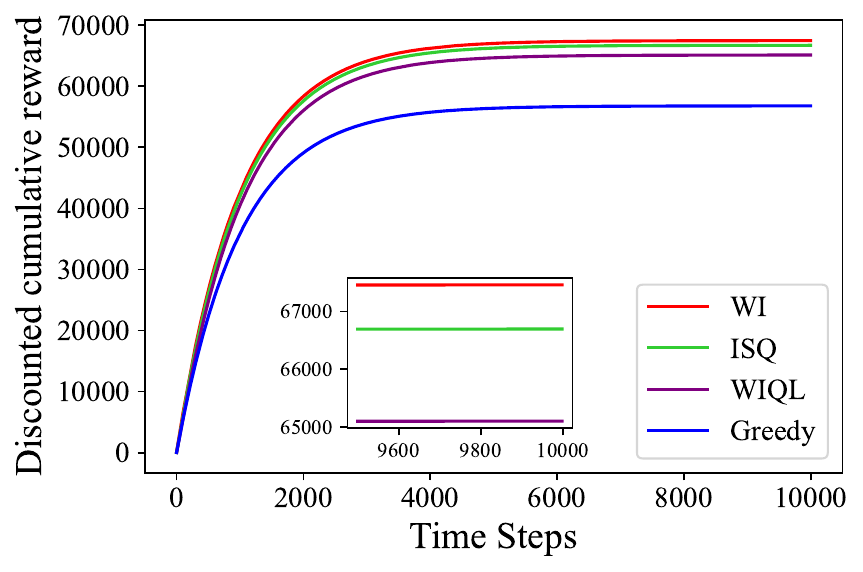}
        \centerline{(b)}
    \end{minipage}
    \caption{Discounted cumulative reward for heterogeneous smart targets. (a) $N=5$, $K=1$. (b) $N=100$, $K=20$.}
    \label{fig3}
\end{figure}

In Fig.~\ref{fig3}, we demonstrate the cumulative discounted rewards of the four policies for $N=5$, $K=1$, and $N=100$, $K=20$.
ISQ always achieves the highest rewards, compared to WIQL and Greedy, and is closest to the upper bound WI.
In particular, ISQ significantly outperforms WIQL and Greedy in all the tested cases.
For the cases with $N=5$ and $N=100$, ISQ improves up to 1.02\% and 2.62\%, respectively, higher rewards than that of WIQL.
The results are consistent with those discussed in Section~\ref{subsec:IVB}.

In all three illustrated cases, especially in the third case, the proposed ISQ policy obtains a faster convergence of Q values, yielding the best learning performance among all the tested policies that do not assume full knowledge of transition matrices.
ISQ also achieves a performance closest to that of WI - the upper bound policy requesting known transition matrices.
The numerical simulation results demonstrate the superiority of the ISQ policy, substantiating its effectiveness and robustness across diverse problem settings.

\section{Conclusion}\label{sec:conclusion}
We have formulated the multi-smart-target tracking problem with active and passive radars as a discrete-time restless multi-armed bandit problem with discrete state space, where the transition matrices of each target are not known a priori.
We have proposed a novel scheduling policy for the radars, the ISQ policy, that incorporates reinforcement learning, i.e., Sarsa and Q-learning, with the index policy.
It enables the APRN to learn the Q values of the underlying bandit processes which are used to approximate the indices - the marginal rewards - of each target.
The indices are being updated along with the controlling process and are used to assist subsequent control decisions through index-based prioritization of the targets.
The efficacy of the proposed ISQ policy has been numerically demonstrated in both the circulant dynamics and smart targets cases. It has been highlighted that ISQ achieves not only higher performance, compared to WIQL, AB, and the Greedy policy, in all the simulated cases but also faster convergence rates with respect to Q values.

{\appendices
\section{The Whittle index policy}\label{AppWI}
Consider a relaxed version of problem \eqref{eq2} that decomposes the original problem into $N$ independent \emph{sub-problems}.
More precisely, the relaxed problem of \eqref{eq2} is
\begin{equation}\label{eq3}
\begin{aligned}
    \max_{\pi} &~\mathrm{E}\left[ \sum_{t=0}^{\infty}\sum_{n=1}^N \beta^t R(X^n_t,a^n_t) \right], \\
    & \text{s.t.}~ \mathrm{E}\left[\sum_{t=0}^{\infty}\sum_{n=1}^N \beta^t a^n_t\right] = \frac{K}{1-\beta},
\end{aligned}
\end{equation}
where the constraint is obtained by taking expectation on both sides of \eqref{eq1}, representing that the expected aggregate discounted number of tracked targets over the infinite time horizon is equal to $K/(1-\beta)$.

We further relax the problem \eqref{eq3} through a Lagrange multiplier $\lambda$, obtaining the dual function of \eqref{eq3},
\begin{equation}\label{eq4}
\begin{aligned}
L(\lambda)=\max_{\pi} ~\mathrm{E}\left[ \sum_{t=0}^{\infty}\sum_{n=1}^N \beta^t\left( R(X^n_t,a^n_t)
+\lambda(1-a^n_t)\right) \right] \\
+ \lambda\left(\frac{K-N}{1-\beta}\right).
\end{aligned}
\end{equation}
Let
\begin{equation}\label{eq4subPro}
L_n(\lambda)=\max_{\pi^n} ~\mathrm{E} \left[\sum_{t=0}^{\infty} \beta^t\left( R(X^n_t,a^n_t) + \lambda(1-a^n_t)\right)\right],
\end{equation}
where $\pi^n$ is an admissible scheduling policy for tracking a single smart target $n$, and we can rewrite \eqref{eq4} as
\begin{equation}
    L(\lambda) = \sum\limits_{n=1}^N L_n(\lambda) + \lambda\left(\frac{K-N}{1-\beta}\right).
\end{equation}

Relaxing the problem \eqref{eq3} to \eqref{eq4} enables us to decompose the combinatorial problem into $N$ independent problems, each of which is associated with a target $n$ and formulated in \eqref{eq4subPro}.

The Whittle index policy is built on the \emph{Whittle indices} that are real numbers assigned to each of the state-target pair. For a given state-target pair, the index intuitively represents the marginal reward of taking action $1$ (actively tracking the target) other than $0$ (passively tracking the target) when the target is in the given state. In other words, for the sub-problem in \eqref{eq4subPro} and a given state $X^n_t$ of target $n$, the index is equal to a special value of $\lambda$, for which $a^n_t=1$ and $0$ are equally attractive.

\section{Indexability}\label{App1}
The \emph{indexability} \cite{whittle1988restless} is the basis of adopting the Whittle index policy and the analysis of its asymptotic optimality~\cite{weber1990index,fu2022restless,fu2020energy,fu2024restless}.
Indexability requires the existence of a threshold-style optimal solution for each sub-problem in~\eqref{eq4}.
For instance, the MTT problems have been proven to be indexable~\cite{nisno2009multitarget,nino2011sensor}.

Base on the definition in~\cite{whittle1988restless}, the Whittle index $\lambda^{n*}: X\to\mathbb{R}$ of target $n$ exists if for any $\lambda\in\mathbb{R}$, it is optimal to take action $a_{n,t}=1~(0)$ when $\lambda^{n*}\ge\lambda$ $\left(\lambda^{n*}<\lambda\right)$.
For the sub-problem associated with target $n$, denote a set of states as $\mathcal{L}(\lambda)$ upon a specific value of $\lambda$, signifying the collection of states wherein it is optimal to activate the corresponding arm/bandit process/target. The indexability of an RMAB problem is given by Definition \ref{d1}.
\begin{definition}[Indexability]\label{d1}
An arm is said to be indexable if $\mathcal{L}(\lambda)$ decreases monotonically from the set of all states to the empty set as $\lambda$ increases from $-\infty$ to $\infty$. A restless bandit problem is said to be indexable if all arms are indexable.
\end{definition}

Based on Definition \ref{d1} and \cite[Definition 3]{nakhleh2021neurwin}, we can define a sufficient condition for indexability, referred to as the \emph{strong indexability}.
For the sub-problem associated with target $n$, we define the expected cumulative discounted reward as $Q_{\lambda}^{act}(X)$, when activating the arm/bandit process/target at time 0 and then employing the optimal policy after time 1.
Conversely, when this policy does not activate the arm at time 0 and employs the optimal policy after time 1, we obtain the cumulative discounted reward $Q_{\lambda}^{pass}(X)$.
Let $D_X(\lambda)\triangleq Q_{\lambda}^{act}(X)-Q_{\lambda}^{pass}(X)$ represent the difference between the rewards of the above-mentioned two policies.
\begin{definition}[Strong Indexability]\label{d2}
An arm is said to be strongly indexable if $D_X(\lambda)$ is strictly decreasing in $\lambda$ for every state $X$.
\end{definition}
}

\bibliographystyle{IEEEtran}
\bibliography{IEEEabrv,reference}

\end{document}